\documentclass[twocolumn,aps,nofootinbib]{revtex4-1}
\usepackage{amsfonts, amssymb}
\usepackage{graphicx, epsfig,bm}
\usepackage{color}

\usepackage{amsbsy}
\usepackage{amsfonts}
\usepackage{amsmath}
\usepackage{amssymb}

\newcommand{\be}{\begin{equation}}
\newcommand{\ee}{\end{equation}}
\newcommand{\bea}{\begin{eqnarray}}
\newcommand{\eea}{\end{eqnarray}}

\def\fun#1#2{\lower3.6pt\vbox{\baselineskip0pt\lineskip.9pt
        \ialign{$\mathsurround=0pt#1\hfill##\hfil$\crcr#2\crcr\sim\crcr}}}

\renewcommand\[{\left[}
\renewcommand\]{\right]}

\newcommand\lsim{\mathrel{\rlap{\lower4pt\hbox{\hskip1pt$\sim$}}
    \raise1pt\hbox{$<$}}}
\newcommand\gsim{\mathrel{\rlap{\lower4pt\hbox{\hskip1pt$\sim$}}
    \raise1pt\hbox{$>$}}}

\def\dslash{\not{\hbox{\kern-2pt $\partial$}}}
\def\Dslash{\not{\hbox{\kern-4pt $D$}}}
\def\Oslash{\not{\hbox{\kern-4pt $O$}}}
\def\Qslash{\not{\hbox{\kern-4pt $Q$}}}
\def\pslash{\not{\hbox{\kern-2.3pt $p$}}}
\def\kslash{\not{\hbox{\kern-2.3pt $k$}}}
\def\qslash{\not{\hbox{\kern-2.3pt $q$}}}

 \newtoks\slashfraction
 \slashfraction={.13}
 \def\slash#1{\setbox0\hbox{$ #1 $}
 \setbox0\hbox to \the\slashfraction\wd0{\hss \box0}/\box0 }

\def\ee{\end{equation}}
\def\be{\begin{equation}}

\newcommand\Tr{{\rm Tr}\,}

\begin{document}
\setlength{\unitlength}{1mm}
\title{Weak lensing and dark energy: the impact of dark energy\\ on nonlinear dark matter clustering}
\author{Shahab Joudaki, Asantha Cooray} 
\affiliation{Center for Cosmology, Dept. of Physics \& Astronomy, University of California, Irvine, CA 92697}
\author{Daniel E. Holz}
\affiliation{Theoretical    Division,   Los   Alamos   National Laboratory, Los  Alamos, NM  87545}

\date{\today}

\begin{abstract}
We examine the influence of percent-level dark energy corrections to
the nonlinear matter power spectrum on constraints of the dark
energy equation of state from future weak lensing probes.
We explicitly show that a poor approximation (off by $\gsim10\%$) to the nonlinear
corrections causes a $\gsim1\sigma$ bias on the determination of the dark
energy equation of state. Future weak lensing surveys must
therefore incorporate dark energy modifications to the nonlinear matter power spectrum
accurate to the percent-level, to avoid introducing significant bias in their
measurements. For the WMAP5 cosmology,
the more accurate power spectrum is more sensitive to dark energy
properties, resulting in a factor of two improvement in dark energy equation of state constraints.
We explore the complementary constraints on dark energy from future weak
lensing and supernova surveys. A space-based, JDEM-like survey
measures the equation of state in five independent redshift bins to $\sim10\%$,
while this improves to $\sim5\%$ for a wide-field ground-based survey like LSST.
These constraints are contingent upon our ability to control weak lensing 
systematic uncertainties to the sub-percent level.
\end{abstract}
\bigskip

\maketitle

\section{Introduction}

The images of distant galaxies are gravitationally lensed by matter
inhomogeneities along the line-of-sight. In the weak lensing regime these
percent-level magnifications and shape distortions of galaxies need to be
analyzed statistically (see~\cite{Refregier,Schneider} for a review). By extracting the shear power spectrum of weakly lensed 
sources \cite{Wittman, KaiserWL, BaconRE, Waerbeke, Jarvis, Hoekstra}, the nature
of the dark energy has been constrained with lensing surveys~\cite{Jarvis, Hoekstra}.

In a comprehensive analysis of future dark energy probes by the Dark
Energy Task Force (DETF), weak lensing is singled out as particularly
promising, in comparison with supernovae (SNe), galaxy cluster counting, and baryon acoustic oscillations (BAOs)~\cite{Albrecht}.
An important aspect is that the lensing power spectrum depends on both
the lensing kernel and the matter power spectrum, making lensing a powerful
probe of both background cosmology and the growth of structure.

The optimism associated with lensing is predicated on overcoming the vast systematic uncertainties in both
measurement and in theory~\cite{Huterer, Dragan, Hirata, Scoccetal, HirSel,
MaHu, White, ZhanKnox, Rudd, Hut06, McDonald, CooHu,ShaCoo}.  These systematics
include dark energy corrections to the modeling of the nonlinear matter
power spectrum~\cite{Huterer, McDonald}, higher order correction terms in the
lensing integral (such as due to the Born approximation and lens-lens coupling~\cite{CooHu,HirSel,ShaCoo}), and uncertainties of the matter
power spectrum on nonlinear scales due to the strong influence of baryonic physics~\cite{Dragan,
ZhanKnox, White, Rudd}. Observational systematics include photometric
redshift uncertainties, multiplicative factors in shear due to calibration
errors, and additive factors due to PSF anisotropies~\cite{MaHu, Hut06}.

Furthermore, the observed ellipticities of weakly lensed galaxies are sensitive
to the reduced shear, $g = \gamma / (1-\kappa)$, where $\gamma$ is the shear and
$\kappa$ is the convergence. In the weak lensing regime we make use of
the expansion of the reduced shear to first order in the fields: $g \approx
\gamma$. For future lensing surveys, this approximation induces a
bias on the cosmological parameters at the same order
as that of the parameter constraints~\cite{Shapiro,DodShaWhi}. For
current purposes, however, we continue to make use of this simplifying assumption of the shear
as the lensing observable.

In this work we examine one particular lensing systematic: dark
energy corrections to the nonlinear matter power spectrum, and the impact of
these on dark energy constraints from the weak lensing power spectrum.
We utilize two approaches towards modeling the nonlinear
matter power spectrum in an evolving equation of state (EOS) environment. Both approaches are based
upon the Smith et al.~(2003)~\cite{Smith} prescription, valid for $w=-1$. The
conventional route in computing the matter power spectrum for $w(z) \neq -1$
has been to use the $w=-1$ fitting functions of Smith et al.~with
appropriate modifications to the growth function and the redshift dependence of
the matter density. 
We compare this method to that developed by McDonald,
Trac, $\&$ Contaldi (2006)~\cite{McDonald} for a constant $w$ cosmology, where numerical simulations
underlie a fitting scheme that provides corrections to the Smith et
al. results. This latter approach approximates the dependence of
the matter power spectrum on a constant dark energy EOS to the level of a few percent. 
We further analyze measurements of the dark energy EOS in (decorrelated) redshift bins, as 
well as direct measurements of a two-parameter Taylor expansion form for the EOS.

Our calculational methods are described in Section~2, with a basic review of the
computation of distances, the growth function, the nonlinear matter power
spectrum, and the weak lensing power spectrum. In
Section 3, we provide the dark energy constraints from weak lensing
tomography, utilizing two different approaches for calculating the nonlinear
matter power spectrum in a $w(z) \neq -1$ cosmology. We examine the
complementarity between future weak lensing surveys and expansion history
probes (e.g. supernovae distances measurements at $z<1.8$). We also
discuss the bias in dark energy due to uncertainties in the nonlinear matter
power spectrum. At the end of Section 3 we provide an exploration of constraint
contamination due to observational systematic uncertainties. Section~4
concludes with a discussion of our findings. We take our fiducial cosmological model, in accordance with WMAP5 data, to
be a flat $\Lambda$CDM universe with $\Omega_c = 0.215$, $\Omega_b = 0.045$, $h = 0.72$, $n_s = 0.96$, $\sigma_8 = 0.8$, and
no massive neutrinos~\cite{Dunkley}.

\section{Calculational Method}

We begin with a summary of our calculation. We briefly describe our cosmological
distance and growth function, and then discuss the relevant observational
quantities for weak lensing.

\subsection{Distances}

The comoving distance to an object at redshift $z$ is
\begin{equation}
\chi(z) = {1 \over {\sqrt{| \Omega_k |} H_0 }} X \left ({\sqrt{| \Omega_k |}  } \int_0^z {dz' \over {H(z')/H_0}} \right) ,
\end{equation}
where $X(x) = \sin(x)$ for a closed universe, $\sinh(x)$ for an open
universe, and $x$ for a flat universe, and where $H_0 = 100 ~ h$ km s$^{-1}$
Mpc$^{-1}$ is the Hubble constant. The Hubble parameter is given by
\begin{equation}
H(z) = H_0 \sqrt{\Omega_m (1+z)^3 + \Omega_w F(z) + \Omega_k (1+z)^2} ,
\end{equation}
where $\{ \Omega_m, \Omega_w, \Omega_k \}$ are the present matter, dark energy,
and curvature densities, in units of the critical density. The evolution of the
dark energy is represented by $F(z)$. For a cosmological constant,
$F(z)\equiv1$.

\begin{figure}[!t]
\epsfxsize=3.4in
\centerline{\epsfbox{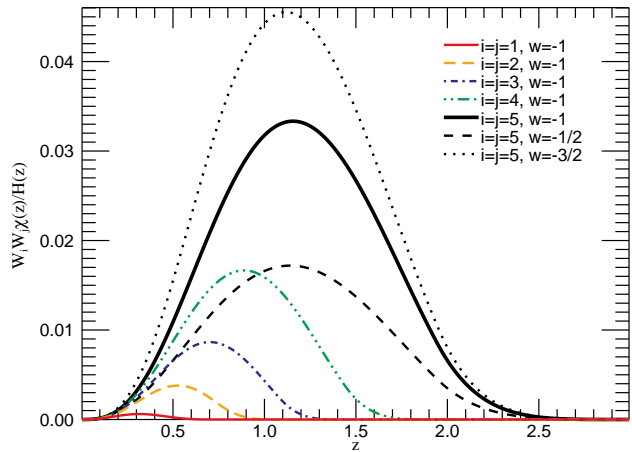}}
\caption{Geometric factor, $W_i(z)W_j(z)\chi(z)/H(z)$, of five tomographic bins
for our fiducial model ($w=-1$). For the fifth bin we also plot the
kernel for $w=-1/2$ and $w=-3/2$. The narrowing of the lensing kernel for $w>-1$ stems
from the decrease of each term with increasing $w$.}
\label{figure1}
\end{figure}

We parameterize the evolution of the dark energy in two distinct ways. We
use a popular Taylor expanded form for the dark energy EOS, given by $w(z) =
w_0 + {z / (1+z)}w_a$~\cite{Linder}, from which one obtains $F(z) =
(1+z)^{3(1+w_0+w_a)} e^{-3w_a {z \over {1+z}}}$.
Alternatively, instead of assuming a particular (physically unmotivated) model
for the EOS, we utilize an agnostic, model-free approach to the redshift
evolution of the dark energy~\cite{HutSta, HuCo, Sullivan, Sarkar}. We fit for $w(z)$ binned in redshift:
\begin{equation}
w(z) = \sum_{i=1}^N w_i \Xi(z_i,z_{i+1}) ,
\end{equation}
where $\Xi(z_i,z_{i+1})$ is a tophat function in the region spanned by $\{
z_i,z_{i+1} \}$, but the ensuing analysis decorrelates the redshift
bins. For this model-independent parameterization of the EOS,
\begin{equation}
F(z_{n-1}<z<z_n) = (1+z)^{3(1+w_n)} \prod_{i=0}^{n-1} (1+z_i)^{3(w_i-w_{i+1})}.
\end{equation}

\subsection{Growth Function}

For a particular parameterization of the EOS one can calculate the growth of
matter fluctuations in the universe. For matter perturbations on linear scales,
it is possible to separate out the time evolution of the perturbation:
$\delta({\bf k},z)/\delta({\bf k},0)=D(z)/D(0)$, where $D(z)$ is the growth function
(which evolves as the scale factor in a matter dominated universe). For $w(z)
\neq -1$ this linear approximation breaks down on very large scales due to
clustering in the dark energy. However, due to the large
uncertainties from cosmic variance at these scales, the impact of dark energy inhomogeneities
are negligible for weak lensing studies (e.g.~\cite{Huterer,SongKnox}).

The normalized growth function, $G(z) = (1+z) D(z)$, can be found by solving a second-order differential
equation~\cite{LinJen}:
\begin{equation}
\begin{split}
G'' + \left[ {7 \over 2} - {3 \over 2} {{w(a) \Omega_w(a)} \over {\Omega_m(a)+\Omega_w(a)}} \right] {G' \over a} ~ + \\
{3 \over 2} \left[ {\Omega_w (1-w(a)) \over {\Omega_m(a)+\Omega_w(a)}} \right] {G \over a^2} = 0 .
\end{split}
\end{equation}
This differential equation is valid for non-flat geometries,
and carries the initial conditions $\{ G(z_{{\rm md}}) = 1; ~ {dG \over
dz}|_{z_{{\rm md}}} =0 \}$ in a matter dominated epoch $z_{{\rm md}}$.

\subsection{Weak Lensing Observables}

We employ weak lensing tomography, wherein we divide the redshift distribution of
source galaxies into distinct redshift bins~\cite{Hu99}. This provides information about the
redshift distribution of the intervening lenses, and thereby allows for more
stringent constraints on cosmological parameters~\cite{Hu99,MaHu}. 

The number density of source galaxies in a square arcminute in each tomographic
redshift bin (with boundaries $z_i<z_s<z_{i+1}$) is
defined by $\bar{n}_i = \int_{z_i}^{z_{i+1}}\! dz_s\,\rho (z_s)$, where
$\rho (z_s) = \bar{n}_g {z^{\alpha} \over 2 z_0^3} e^{-(z_s/z_0)^{\beta}}$ is
the redshift distribution of source galaxies~\cite{Wittman}. We adopt $\left\{
z_0=0.5,\ \alpha=2,\ \beta=1 \right\}$~\cite{Zhan}, appropriate for the Large
Synoptic Survey Telescope~(LSST~\cite{LSSTsite, Ivezic}), normalized such that
$\int_0^\infty dz\,\rho (z) = \bar{n}_g$. We use the same distribution
to describe our Joint Dark Energy Mission (JDEM) source population, but with improved $\bar{n}_g$
(see Table~~\ref{table1}).
We use the Born approximation, and perform the lensing calculation along the unperturbed 
photon path~\cite{CooHu,HirSel,ShaCoo}. The lensing weight function of the $i^{th}$ tomographic
bin is given by
\begin{eqnarray}
\lefteqn{W_i(z) = {3 \over 2} {\Omega_{m} H_0^2 \over \bar{n}_i} (1+z)\chi(z) ~ \times}\\
&&\hspace*{1cm}\int_{\max(z,z_i)}^{z_{i+1}} dz_s {\chi(x(z_s)-x(z)) \over \chi(z_s)} \rho(z_s) ,\nonumber
\end{eqnarray}
for $z \leq z_{i+1}$, where $x(z)={\sqrt{| \Omega_k |}  } \int_0^z {dz' \over
{H(z')/H_0}}$. To ensure that we have no lenses behind our sources, we 
take $W_i(z) = 0$ for $z>z_{i+1}$. The weight function increases for a more
negative dark energy equation of state, as shown in Figure~\ref{figure1}.
The power spectrum of the convergence field is subsequently given by the Limber approximation~\cite{Limber}:
\begin{equation}
C_{ij}(l) = {2\pi^2 \over l^3}\int_0^{z_{\rm H}}dz{W_{i}(z)W_{j}(z)\chi(z) \over H(z)}\Delta_{{\rm NL}}^2({l/\chi(z)},z) ,
\end{equation}
where $z_{\rm H}$ is the horizon redshift, and $\Delta_{{\rm NL}}^2({l/\chi(z)},z)$ is the full nonlinear matter power spectrum.

For $l \gsim 300$, the dominant contribution to the matter power spectrum will be from nonlinear 
scales~\cite{JainSeljak,Cooray,TakadaJain}, emphasizing the need to correctly model the
effect of dark energy at these scales. The observed convergence power spectrum,
which is identical to that of the shear~\cite{Schneider}, is contaminated
by shot noise due to the finite source density, as well as uncertainty in the
intrinsic shapes of the source galaxies, leading to: $\tilde{C}_{ij}(l) =
C_{ij}(l) + \delta_{ij} \left\langle \gamma^2
\right\rangle /\bar{n}_i$ (which assumes that the noise is uncorrelated between
tomographic bins). Whereas cosmic variance dominates the error
on large angular scales, the shot noise is dominant on small scales. 

For simplicity we take the intrinsic shape uncertainty of the source galaxies to
be redshift independent: $\left\langle \gamma^2 \right\rangle ^{1/2} = 0.22$, in
accordance with expected results from a future ground-based survey such as
LSST~\cite{Zhan,LSSTsite}. We assume that the survey
covers half of the sky, with a galaxy density of $\bar{n}_g = 50$ arcmin$^{-2}$~\cite{Zhan, LSSTsite}. For 
comparison, we also consider a space-based survey,
such as a JDEM candidate like the SuperNova
Acceleration Probe (SNAP)~\cite{SNAPsite, Aldering, RefregSNAP}. For simplicity,
we keep the same source distribution and intrinsic shear uncertainty, modifying
the source density to twice that of the ground-based survey,
and the width of the survey to a tenth of the sky. The characteristics of the two
surveys are summarized in Table~\ref{table1}.

\begin{figure}[!t]
\epsfxsize=3.4in
\centerline{\epsfbox{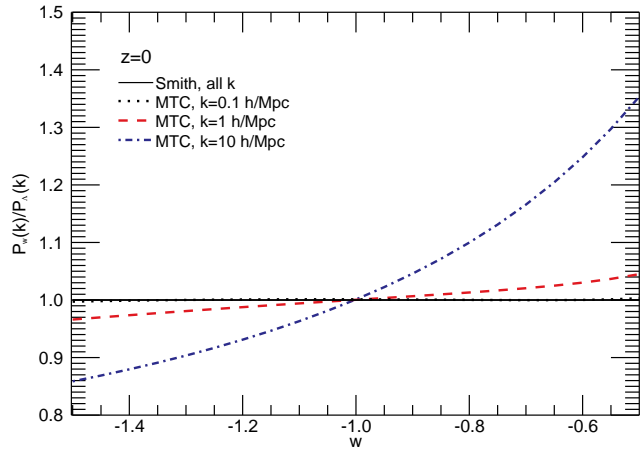}}
\caption{Ratio of the present matter power spectrum of a constant dark
energy EOS to that of a cosmological constant,
$\Delta^2_w(k,z)/\Delta^2_\Lambda(k,z)$, as a function of $w$.
The solid line is obtained by modifying the growth
function and matter density evolution based on the Smith et
al. prescription, while the other (more accurate) curves utilize the MTC
corrections to the power spectrum.}
\label{figure2}
\end{figure}

\subsection{Nonlinear Matter Power Spectrum}

We now detail our calculation of the matter power spectrum for a general dark
energy cosmology.\footnote{We do not pursue a direct halo-model
approach~\cite{CooraySheth} for modeling the weak lensing power spectrum, since
for the case of general dark energy models we have inadequate descriptions for
the halo mass function, the halo dark matter profile, and the large-scale halo
bias.}  In the linear regime the transfer function $T(k,z)$ is computed using
the prescription of Eisenstein $\&$ Hu (1997)~\cite{EisHu}. The dimensionless
linear power spectrum, normalized to the variance of the matter density
field on scales of $8 ~ h^{-1}$ Mpc, $\sigma_8$, is given by
\begin{equation}
\Delta_{\rm L}^2(k,z) =  {k^{3+n_s} T^2(k,z) {D^2(z) \over D^2(0)}} \left({\sigma_{8,{\rm obs}} \over \sigma_{8,{\rm theory}}}\right)^2 ,
\end{equation}
where $n_s$ is the spectral index, $\sigma_{8,{\rm obs}} = 0.8$, and
\begin{equation}
\sigma_{8,{\rm theory}} = \sqrt{\int d\ln k \Delta_L^2(k,0) J^2(8 k')}\ ,
\end{equation}
with the spherical tophat filter, $J(8k') = [3/(8k')^3][\sin(8k') -
(8k')\cos(8k')]$, and $k'=k$ $h^{-1}$ Mpc is dimensionless. One can subsequently
extend this power spectrum to nonlinear scales by calculating the appropriate effective
spectral index, effective spectral curvature, and nonlinear scale, employing
the fitting functions provided in Smith et al.~\cite{Smith}.

\begin{figure}[!t]
\epsfxsize=3.4in
\centerline{\epsfbox{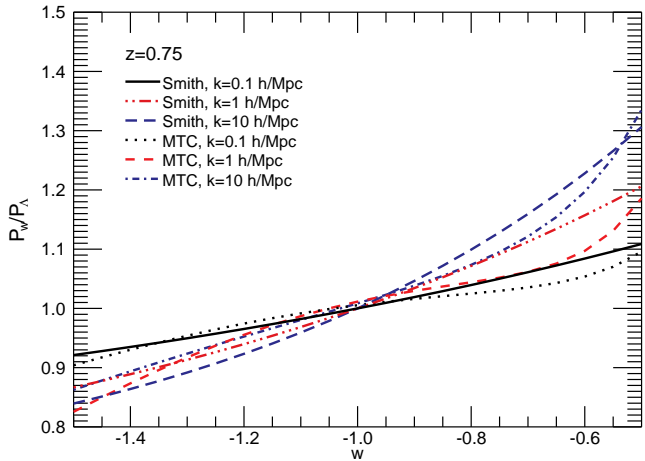}}
\caption{Ratio of the matter power spectrum of a constant dark energy EOS to
that of a cosmological constant, at z=0.75.}
\label{figure3}
\end{figure}

\begin{table}[!b]\footnotesize
\begin{tabular}{r|cccccccccc|c}
\hline
Probe  & $f_{\rm {sky}}$ & $\bar{n}_g$ ({arcmin}$^{-2}$) & $z_{\rm {peak}}$ & $\sqrt{\left\langle \gamma^2 \right\rangle}$ \\
\hline
LSST   &  0.5       & 50   & 1.0     &  0.22  \\
\hline
{JDEM}   &  0.1       & 100   & 1.0     &  0.22 \\
\end{tabular}
\caption{Descriptions of our fiducial ground-based (LSST) and space-based (JDEM) probes.}
\label{table1}
\end{table}
The underlying cosmology in the Smith et al.~fitting functions manifests itself
in two distinct ways. First, cosmology impacts the evolution of the matter
density, $\Omega_m(z)$, and the evolution of the growth of matter perturbations,
$D(z)$. Second, cosmology fixes the functional form and coefficients associated
with the fitting functions, which are fine-tuned to a suite of $\Lambda$CDM
N-body simulations. Thus, whereas an arbitrary dark energy EOS could make an
imprint on the matter power spectrum via its influence on $\Omega_m(z)$ and
$D(z)$, the cosmological dependence of the N-body fitting functions remain
fine-tuned to a $w\equiv-1$ dark energy EOS.
A correct approximation of the matter power spectrum in an evolving EOS
cosmology therefore requires both an update of the effective spectral
parameters of the Smith et al.~prescription, and a generalization of the fitting 
functions.

\begin{table}[htb]\footnotesize
\begin{tabular}{r|cccccccccc|c}\hline
No. of bins & $z_1$ & $z_2$ & $z_3$& $z_4$ & $z_5$ \\
\hline
1 bin    & $3.0$       & ---       & ---       & --- & --- \\
\hline
2 bins    & $1.3$       & $3.0$       & ---       & --- & --- \\
\hline
3 bins    & $1.0$       & $1.6$       & $3.0$       & --- & --- \\
\hline
4 bins    & $0.85 $      & $1.3 $      & $1.85 $      & $3.0$ & --- \\
\hline
5 bins    & $0.75 $      & $1.1 $      & $1.45 $   &  $1.95$ &  $3.0$ \\
\end{tabular}
\caption{Redshift bins of the source distribution. The redshifts are determined
such that each tomographic bin contains 
roughly the same number density of galaxies. We ignore sources above $z=3$.}
\label{table2}
\end{table}

\begin{figure}[!b]
\epsfxsize=3.4in
\centerline{\epsfbox{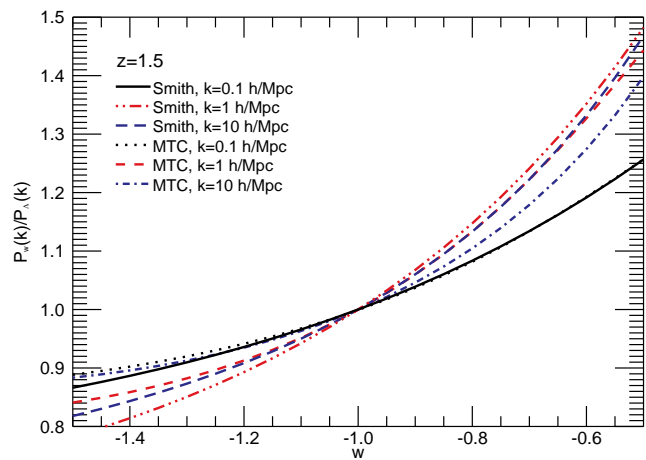}}
\caption{Ratio of the matter power spectrum of a constant dark energy EOS to
that of a cosmological constant, at z=1.5.}
\label{figure4}
\end{figure}

A full calculation of the matter power spectrum will thus require a large suite
of N-body simulations. An alternative approach devised by McDonald, Trac, $\&$
Contaldi (2006; hereafter MTC) notes that many numerical uncertainties cancel
when taking the ratios of power spectra~\cite{McDonald}. By interpolating the
matter power spectrum ratios between constant EOS values, MTC provide a route
to a fast calculation of the nonlinear matter power spectrum in constant $w$
cosmologies. An accurate matter power spectrum can thereby be obtained by
computing the power spectrum for a $\Lambda$CDM universe, and subsequently
multiplying it by a cosmology-dependent correction factor. In particular,
\begin{eqnarray}
\lefteqn{\Delta^2_{{\rm NL},w}(k,z)=\Delta^2_{{\rm NL},\Lambda}(k,z)\times{}}\\
&&{{D^2(z,w)/D^2(0,w)} \over {D^2(z,\bar w)/D^2(0,\bar w)}}
\Upsilon(k,z,{\bf p})\ ,\nonumber
\end{eqnarray}
where the $N$ cosmological parameters are given by ${\bf p} = \{ \Omega_m, \Omega_b, h, 
\sigma_8, n_s, w\} $, $\bar w\equiv-1$, and
\begin{equation}
\Upsilon(k,z,{\bf p}) = e^{{\left(\prod_{i=1}^N \sum_{\nu_i=0}^M p_i^{\nu_i}\right) A_{{\nu_1}{\nu_2}...{\nu_N}}(k)}}.
\end{equation}
In this equation, $M$ is the polynomial order of the parameters, and the
coefficients $A_{{\nu_1}{\nu_2}...{\nu_N}}(k)$ are provided by a least squares
fit to the MTC simulations. On large scales the correction factor is equal to
unity, as expected. The power spectrum ratios are only provided up to $z=3/2$
and $k=10 ~ h$ Mpc$^{-1}$. 
However, since we only consider multipoles up to $l=3000$, scales smaller than $k=10 ~ h$
Mpc$^{-1}$ are only probed at very low redshift, where the number density of sources becomes
negligible.
Moreover, as gradually smaller scales become linear with increasing redshift, for a fixed range of scales, the influence of dark energy on the matter power spectrum is progressively modelled by linear theory as the redshift increases.
The dark energy correction factor for the matter power spectrum therefore covers an adequate range 
of scales and redshifts for weak lensing studies.
Furthermore, as the MTC simulations have only been carried out for a limited range of parameter space, we have ensured that all 
of the cosmological parameters under consideration here live within their explored region.

\begin{figure}[!t]
\epsfxsize=3.4in
\centerline{\epsfbox{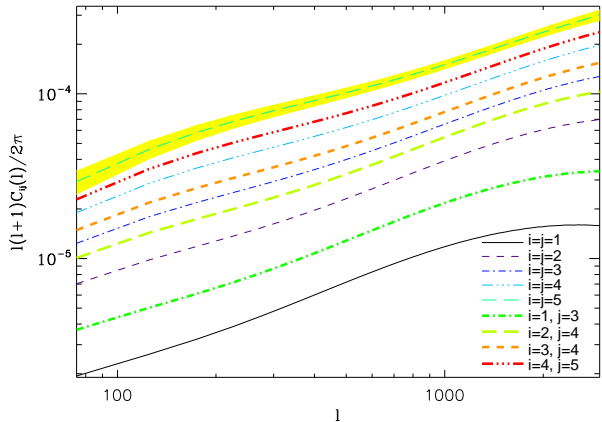}}
\caption{Convergence power spectra for the case of five tomographic bins in a $\Lambda$CDM cosmology. We
also include a number of representative cross-terms. For the $i=j=5$ case, we include
the noise contribution as a band about the curve.}
\label{figure5}
\end{figure}

\section{Results}

We next explore the differences between the Smith et al.~and MTC matter power spectra, and their
respective impacts on cosmological constraints from future weak lensing
probes. Systematic uncertainties are discussed in Section~G.

\subsection{Matter Power Spectrum: MTC vs. Smith}

In Figures~2--4 we show the evolution of the MTC nonlinear matter power spectrum as a
function of dark energy EOS (for constant $w$). Each figure represents a
different redshift ($z = 0$, 0.75, and 1.5), and every plot shows three
different MTC curves, corresponding to scales of $k = \[0.1, 1.0, 10 \]$ $h$
Mpc$^{-1}$. The curves are normalized to the power spectrum corresponding to a
pure cosmological constant, and the Smith et al.~fit (modifying solely the growth
function and the matter density in accordance with the dark energy EOS) is
also shown, for comparison. The MTC power spectra show large ($\gsim10$\%)
deviations from the simplified Smith et al.~approximation over a range of reasonable
$w$ values. In general, constant $w<-1$ models lead to a suppression of the matter power
spectrum as compared to that in a $\Lambda$CDM universe, with the suppresion
more pronounced at smaller (nonlinear) scales. Conversely, the matter power spectrum is
enhanced for $w > -1$. As expected, the MTC correction factor becomes less pronounced 
at higher redshift as increasingly smaller scales become linear.

\begin{figure}[!b]
\epsfxsize=3.4in
\centerline{\epsfbox{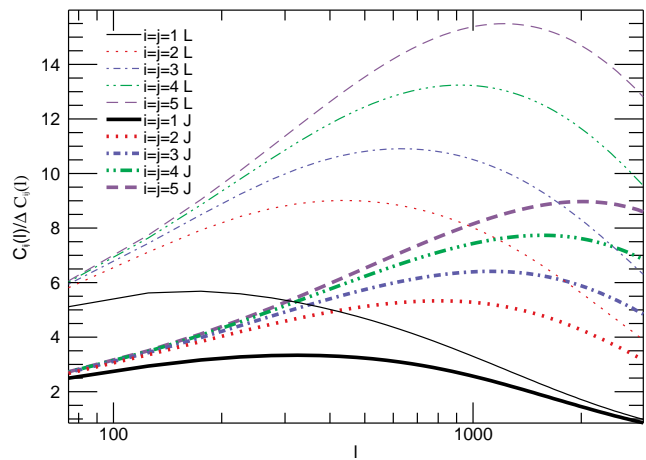}}
\caption{Signal to noise with five tomographic bins in a $\Lambda$CDM
cosmology, for LSST (denoted by `L') and JDEM (`J').}
\label{figure6}
\end{figure}

\subsection{Weak Lensing}

The convergence power spectrum depends on both the geometric factor,
$W_i(z)W_j(z)\chi(z)/H(z)$, and the matter power spectrum, $\Delta^2(k,z)$, but
in a competing manner. Whereas a decreasing dark energy EOS has the effect of
increasing the geometric factor, and therefore the convergence power spectrum,
it also suppresses the nonlinear matter power spectrum, which in turn suppresses
the convergence power spectrum. As a result, these two effects partially cancel,
thereby decreasing the sensitivity of weak lensing to dark energy.

In Figure~\ref{figure5} we plot the power spectrum of the convergence for five
tomographic bins (with the redshift divisions listed in Table~\ref{table2}), as
well as some representative convergence power spectra cross-terms (e.g., between
bins 4 and 5). We divide the power spectrum by its noise, $\Delta
C_{ij}(l) = {f^{-1/2}_{\rm{sky}}} \sqrt{2/(2l+1)} (C_{ij}(l) + \delta_{ij}
\left\langle \gamma^2 \right\rangle /\bar{n}_i)$, in Figure~\ref{figure6}. The
signal to noise is consistently higher for the wider LSST than it is for the
deeper JDEM. The larger width of LSST gives strong signal to noise in
particular at smaller multipoles, while the greater depth of a JDEM-type survey
makes it increasingly competitive at larger multipoles.

\begin{figure}[!t]
\epsfxsize=3.4in
\centerline{\epsfbox{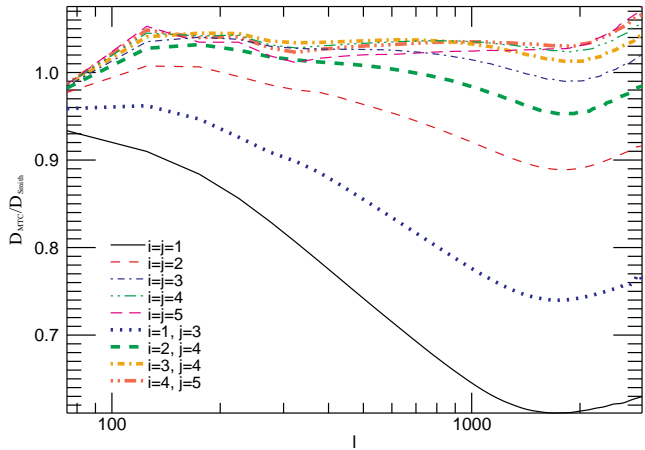}}
\caption{Ratio of MTC to Smith et al.~convergence spectra derivatives with respect to
$w$, i.e. $\rm{D_{MTC}/D_{Smith}} = {{\partial {\bf C}_l} \over {\partial w}} |_{\rm{MTC}} / {{\partial
{\bf C}_l} \over {\partial w}} |_{\rm{Smith}}$.}
\label{figure7}
\end{figure}

Figure~\ref{figure7} presents the ratio of the convergence spectra derivatives
with respect to a constant EOS between Smith et al.~and MTC. The MTC derivatives 
approach the Smith et al.~derivatives in higher-redshift
tomographic bins, as the corresponding matter power spectra converge at large
redshift. This is because the MTC correction factor approaches unity at high~$z$
(see Figs.~2--4), and we therefore impose the
approximation $\Delta^2(k,z) |_{\rm{MTC}} \equiv \Delta^2(k,z)|_{\rm{Smith}}$
for $z>1.5$. Tables~\ref{table3} and~\ref{table5} show that our conclusions are
insensitive to this approximation.
The $z<1.5$ entries in these tables are for five tomographic bins, with redshift
divisions at [0.59,0.83,1.04,1.26,1.5], engineered to preserve an equal number density of
sources across each bin.
Since the MTC derivatives have decreased
relative to the Smith et al.~ones, one might conclude that the sensitivity to dark energy
has also decreased. However, due to the particular nature of the
cross-correlations between cosmological parameters, this turns out not to be the
case for our fiducial cosmology, as is shown below.

\subsection{Constraints on Dark Energy}

In the previous subsections we have examined the corrections to the nonlinear matter
power spectrum due to the presence of dark energy. We now explore how these corrections impact weak lensing
constraints of the dark energy. Assuming Gaussianity in the likelihood distribution of the cosmological parameters for the lensing 
power spectrum \cite{Scoccetal,CoorayHu,WhiteHu1,WhiteHu2,TakadaJain08}, we utilize a Fisher matrix analysis~\cite{HuJain,Tegmark}:
\begin{equation}
F_{\alpha \beta} = f_{\rm sky} \sum_l {(2l+1)\Delta l \over 2} \Tr \left
[{{\partial {\bf C}_l} \over {\partial p_\alpha}} \tilde{{\bf C}}_l^{-1}
{{\partial {\bf C}_l} \over {\partial p_\beta}} \tilde{{\bf C}}_l^{-1} \right ].
\label{eq:fisher}
\end{equation}
We consider constraints for both a ground-based survey, such as LSST, and 
a space-based JDEM probe, such as SNAP. We
analyze multipoles between 50 and 3000, with the power spectrum of the convergence field divided
into multipole bins of width $\Delta l = 50$, (the results are insensitive to
the choice of binning). The cutoff
at $l=3000$ avoids non-Gaussianities of the convergence field~\cite{Scoccetal,CoorayHu,WhiteHu1,WhiteHu2,TakadaJain08}, as well as 
uncertainties from baryonic physics that increase at larger multipoles~\cite{White,ZhanKnox}.

We present constraints on cosmology for up to five tomographic bins, where
each bin is constructed to contain the same effective number density of
galaxies. The redshift divisions of the bins are listed in Table~\ref{table2}.
For the terms in Eq.~\ref{eq:fisher} we carry out two-sided numerical derivatives
with steps of $10 \%$ in parameter value (except for $\pm 0.1$ in the case of $w_a$, and
$\pm 0.05$ for $\Omega_k$). Our results are essentially unaffected by the
particular choice of step size.
We note that there is an ambiguity in the evaluation of the $\Omega_k$
derivative at $\Omega_k=0$, since the Smith et al.~fitting functions are only given for flat and
open universes. Although it is common to carry out the curvature derivative
using the flat cosmology fitting functions, we also evaluate the derivative
at a slightly open universe to see if it renders noticeable differences in
the constraints. This assumes that the
power spectrum doesn't vary sharply about flatness. The two methods for
calculating the $\Omega_k$ derivative yield consistent results (within $\sim15$\%), and we quote
results utilizing the latter method.

\begin{table}[t!]\footnotesize
\begin{center}
\begin{tabular}{r|cccccccccc|c}
\hline
Probe  & $\sigma (w_0)$ & $\sigma (w_a)$ & FOM$_{\rm{DETF}}$ & $\sigma (w_{c})$& FOM$(w_c)$ \\
\hline \hline
1 (M-LSST)     &  0.51     &   1.7       & 1.7  & 0.34 & 8.9 \\
\hline
2 (M-LSST)      & 0.14     &   0.46   &      80 & 0.027   & 1300  \\
\hline
3 (M-LSST)     &  0.082      & 0.31     &   190  & 0.017 & 3500 \\
\hline
4 (M-LSST)     &   0.071      & 0.27 &  240   & 0.0152 &  4300 \\
\hline
5 (M-LSST)     &  0.069      &  0.27   &  250 &  0.0148      &   4600   \\
$+ \Omega_k$     &  0.15      &  0.64 & 96 &    0.0162      &   3800   \\
$+$ Priors     &  0.072      &  0.28 & 230  &  0.0150      &   4500   \\
\hline
$l<1000$     &  0.19      &  0.81   &  57 &  0.022      &   2100   \\
$+ \Omega_k$     &  0.23      &  1.04 & 36  &  0.027      &   1400   \\
$+$ Priors     &  0.14      &  0.59 & 79   &  0.022      &   2200   \\
\hline
$z<1.5$     &   0.099     &  0.37   &  150 &  0.019      &   2900   \\
$+ \Omega_k$     &  0.14      &  0.74 & 32  &  0.043      &   550   \\
$+$ Priors     &  0.010      &  0.37 & 140   &  0.019      &   2700   \\
\hline
$+$ 2300 SNe    &   0.043  & 0.17 &   410   & 0.014  &  4900  \\
\hline
1 (S-LSST)     &  1.4      &  12          &   0.16 &  0.50  & 3.9 \\
\hline
2 (S-LSST)      &  0.20     &  0.50        &   59 &   0.034 & 860 \\
\hline
3 (S-LSST)     &    0.082    & 0.24   & 170 &  0.025 & 1700  \\
\hline
4 (S-LSST)     &  0.069    &  0.21   & 220 &   0.022  & 2100 \\
\hline
5 (S-LSST)     &   0.067     &  0.21 &   230   &  0.021  &  2200  \\
$z<1.5$     &   0.094     &  0.29 &  140  & 0.025  & 1600  \\
\hline
2 (M-JDEM)      & 0.22     &  0.71         & 31  &  0.046 & 480 \\
\hline
5 (M-JDEM)     &   0.11     &  0.44 &  100  & 0.023   & 1900  \\
$+ \Omega_k$     &  0.22      &  0.96 & 42   &  0.025      &   1700   \\
$+$ Priors     &  0.11    &  0.44 & 98  &  0.023      &   1900   \\
\hline
$l<1000$     &   0.30     &  1.3 &  22  & 0.036   & 760  \\
$+ \Omega_k$     &  0.37      &  1.6 & 14 &   0.044   &   530   \\
$+$ Priors     &  0.23    &  0.95 & 29   &  0.036      &  780   \\
\hline
$z<1.5$     &   0.15     &  0.56 &  58  & 0.031   & 1000  \\
$+ \Omega_k$     &  0.20      &  1.1 & 15 &   0.061   &   270   \\
$+$ Priors     &  0.015      &  0.56 & 57   &  0.031      &   1000   \\
\hline
$+$ 2300 SNe    &     0.048 &  0.21 & 230   & 0.021 & 2200    \\
\hline
2 (S-JDEM)      &  0.32     &   0.79       &    21 &  0.059  & 280 \\
\hline
5 (S-JDEM)     &  0.11      &  0.34 &   87  &  0.034     & 860   \\
$z<1.5$     &   0.14     &  0.44 &  54  & 0.041   & 580  \\
\end{tabular}
\caption{FOM values and 1$\sigma$ constraints on the dark energy
parameters, based on either the Smith et al.~prescription (denoted by `S') or 
with the inclusion of MTC corrections (denoted by `M'). The $+2300$ entries in the table are a joint analysis of WL and SNe, with SN systematics included (flatness assumed). The 2300 SNe alone constrain [$w_0,w_a,w_c$] to [0.04, 0.67, 0.036] in the case of statistical noise only, and to [0.055, 1.0, 0.054] when systematic errors are included. For the cases with priors, we utilize an HST prior of 0.08 on $h$~\cite{Freedman} and a Planck prior of 0.0032 on $\Omega_k$~\cite{SmithHuKap}. The strongest improvement comes from the curvature prior.}
\label{table3}
\end{center}
\end{table}

We consider two dark energy models: $w(z) = w_0 + {z / (1+z)}w_a$, and $w(z)=w_c$ is constant (i.e., $w_a=0$). For the
$w_c$ case, we define the figure of merit (FOM) of the EOS measurement to
be ${\rm FOM}(w_c)= \sigma^{-2}(w_c)$. 
For the $\{w_0,w_a \}$ case, we take the FOM to be the inverse 
of the error ellipse in the $\{w_0,w_a \}$ plane (as defined in the DETF report \cite{Albrecht}):
\begin{equation}
{\rm FOM_{DETF}} =  [\sigma(w_p) \sigma(w_a)]^{-1} ,
\end{equation}
where $w_p$ is the pivot value of the EOS.
We do not include the factor of $1/\pi$ in the DETF FOM \cite{Fomswg}.
Note that the uncertainty on the dark energy EOS at the pivot is equal to the uncertainty 
on a constant EOS~\cite{Albrecht}.
We find that there is roughly a factor of two improvement in ${\rm FOM}(w_c)$ in going from
Smith et al.~to the MTC form for the dark matter power spectrum, as shown in Table~\ref{table3}. This relative
improvement in the EOS constraint is independent of the number of tomographic
bins. Although the derivatives in Eq.~\ref{eq:fisher} are larger for the Smith et al.~case (see Fig.~\ref{figure7}), the 
MTC form for the power spectrum provides more stringent constraints due to its reduced cross-correlations in
the Fisher matrix. For a $2\times2$ matrix, this can be visualized as rendering a
larger determinant, and thereby better constraints as the inverse of the Fisher
matrix is taken. Thus, even though smaller derivatives commonly provide poorer
constraints, here we find that the cross-correlation terms
yield improved constraints for a constant (as well as a redshift-binned) EOS.

Generalizing to non-flat cosmologies significantly degrades the cosmological
constraints, especially if we only consider effects out to $z=1.5$.
This degradation is ameliorated with an expected Planck prior of 0.0032 on $\Omega_k$~\cite{SmithHuKap} and 
HST prior of 0.08 on $h$~\cite{Freedman}. For LSST, using the Smith et al.~prescription beyond $z=1.5$, the inclusion of curvature 
drives the FOM down by one sixth from 4600 to 3800, whereas in the case of a
redshift cutoff (at $z=1.5$) the inclusion of the curvature parameter causes a factor of five
deterioration of the dark energy FOM from 2900 to 550. By including HST and
Planck priors, the FOM increases to 4500 and 2700, respectively. A
similar sensitivity to the curvature density is seen for the JDEM case, which 
is also effectively removed by our choice of priors. We moreover consider a lower 
cutoff in $l$ at 1000, noting that this cutoff alone would diminish the 
constraints by roughly a factor of two for LSST and 2.5 for JDEM. 
The relative constraint degradation is larger for a JDEM probe as its integrated signal-to-noise is more
sensitive to larger multipoles (as seen in Fig.~\ref{figure6}).


\subsection{Evolving Dark Energy EOS}

One major limitation of the MTC correction to the nonlinear dark matter power
spectrum is that it is only valid in the case of a constant dark energy EOS.  We
now generalize this approach to the case of dynamical dark energy. We utilize an
observation by Francis et al.~\cite{Francis}, who have shown that the power
spectrum of a dynamical $w(z)$ resembles that of a constant $w$ with the same
distance to the last scattering surface. In other words, when the integrated
expansion history between a dynamical EOS and a constant EOS are similar, the
growth histories will likewise be similar. Francis et al. have numerically
confirmed this to a few percent, at scales up to $k=5$ $h$ Mpc$^{-1}$ for the
$\{w_0,w_a \}$ parameterization. Based upon this, we mimic the matter power
spectrum of an evolving dark energy EOS by utilizing a constant EOS with a
matched distance to the last scattering surface. We calculate all other
components of the convergence spectrum based upon the actual $w(z)$.

The FOM of the DETF remains relatively unchanged despite the inclusion of the improved (MTC) nonlinear matter power spectrum. 
For the $\{w_0,w_a \}$ parameterization, we note that although the constraints due to changes in $w_0$ are improved for MTC, the Smith et al.~constraints are more sensitive to changes in $w_a$, and
therefore their net effects on the FOM are roughly equivalent.
This latter decrease in sensitivity for MTC occurs as the
distance-matching prescription of mapping $\{w_0=-1,w_a\}$ onto $w_c$ renders larger
deviations in the MTC matter power spectrum, and thereby smaller deviations in the
convergence spectrum due to the cancellation with the lensing kernel discussed above. The
smaller $w_a$ derivatives then lead to inferior
constraints. Imposing a cutoff at $l=1000$ renders a factor of five deterioration in the
FOM of the DETF, primarily due to the factor of three deterioration of the
constraint on $w_a$, as shown in Table~\ref{table3}.

In addition to the two-parameter $\{w_0,w_a\}$ constraints discussed above, we also consider
constraints on a redshift-binned EOS. We utilize the MTC method, generalized
following the Francis et al. approach of matching the last-scattering
distances between evolving and non-evolving dark energy. Once the weak
lensing Fisher matrix for correlated dark energy bins is obtained, we rotate the
dark energy parameters into a basis where they are uncorrelated.

The most straightforward approach would be to diagonalize the marginalized Fisher matrix $\tilde{{\bf F}}$
(i.e. marginalized over all parameters except for the EOS, $w_{i=1...{\rm N}}$),
such that the uncertainties in the uncorrelated dark energy parameters are given
by the inverse of the eigenvalues. However, we choose the transformation matrix
advocated in Huterer $\&$ Cooray (2005)~\cite{HuCo}, namely ${\tilde{{\bf F}}}^{1/2} =
{\bf O}^T {\bf \Lambda}^{1/2} {\bf O}$, where ${\bf O}$ and ${\bf \Lambda}$ are
the eigenvector and eigenvalue matrices. The uncorrelated parameters are then
given by ${\bf q} = {\tilde{{\bf F}}}^{1/2} \tilde{{\bf p}}$~\cite{HutSta,HamTeg,HuCo,Sarkar}. This transformation matrix 
has the pleasing feature of generating relatively localized and mostly positive weights~\cite{HamTeg,HuCo}, while 
also preserving the information content of the correlated Fisher matrix. Note 
that the transformation matrix is normalized such that the sum of the elements
along each row is equal to unity~\cite{HamTeg,HuCo}. The rows therefore
represent the contribution of each redshift bin to the decorrelated parameters. 
The errors on the uncorrelated parameters are then given by
\begin{equation}
\left\langle \Delta q_i \Delta q_j \right\rangle = \delta_{ij} \left[ \sum_{\alpha} {\tilde{F}}^{1/2}_{i \alpha} \sum_{\beta} {\tilde{F}}^{1/2}_{j \beta} \right]^{-1}.
\end{equation}
We now define the figure of merit for this parameterization to be that
of Sullivan, Cooray, $\&$ Holz (2007)~\cite{Sullivan}:
\begin{equation}
{\rm FOM}_{{\rm wbinned}} =  \sum_{i} {\sigma^{-2}(w_i)} = {\sigma^{-2}(w_c)} .
\end{equation}
We note that this FOM is equal to that of a constant EOS \cite{LinPut}, which is an
important feature as it explicitly demonstrates that the FOM is binning
independent. This can be seen by comparing the case of seven bins,
$w_{i=1...7}$, with that of one bin, $w_c$, in Tables~\ref{table3} and~\ref{table5}.
In addition, as the parameter constraints scale as $f_{\rm{sky}}^{-1/2}$, both $\rm{FOM_{wbinned}}$ and $\rm{FOM_{DETF}}$ increase linearly with the fraction of the sky covered.

For comparison with the EOS constraints in Sarkar et al., we have chosen the
same redshift binning of the EOS (boundaries at $z = \{ 0.07, 0.15, 0.3, 0.6, 1.2, 3.0\}$). Although 
the seventh bin is highly unconstrained, the uncertainty does not leak noticeably 
into the other parameters (as compared to fixing $w_7 = -1$).
This particular binning attempted to maximize the number
of redshift bins with 10$\%$ or better constraints on the dark energy EOS from future 
cosmic microwave background (CMB), BAO, and SN
data. We have not provided a comparable binning for the case of weak
lensing. One possible approach is to pick redshifts such that the
sensitivity of the integrand to changes in $w$ are equivalent in each bin.

\begin{table}[!t]\footnotesize
\begin{center}
\begin{tabular}{r|cccccccccc|c}
\hline
$p_i$  & (M-LSST) & (M-JDEM) & (S-LSST) & (S-JDEM) \\
\hline \hline
$w_c$  &  0.015 &  0.023  & 0.021    &  0.034     \\
\hline
FOM    &   4600    &  1900    &  2200    &  860   \\
\hline
$\Omega_{c}$ & 0.015    &  0.028 &  0.015  &   0.029   \\
\hline
$\Omega_{b}$  & 0.015   &   0.029    &  0.015  &  0.029      \\
\hline
$h$      &  0.11     &  0.20  &   0.11  &   0.20 \\
\hline
$n$     &   0.016     &   0.030 &    0.016 &  0.031  \\
\hline
$\sigma_{8}$  & 0.0017  &   0.0028    &  0.0019    &  0.0032  \\
\hline \hline
$w_0$      &  0.069     &  0.11  &   0.067  &    0.11   \\
\hline
$w_a$      &  0.27     &  0.44  &   0.21  &   0.34    \\
\hline
FOM    & 250    &   100  &  230   &  87   \\
\hline
$\Omega_{c}$ & 0.015     &   0.030    & 0.016   &  0.029   \\
\hline
$\Omega_{b}$ & 0.015    &   0.029   &  0.015  &  0.029  \\
\hline
$h$     &  0.11     &  0.20  &  0.11   &  0.20 \\
\hline
$n$      &  0.019     &  0.035  &  0.016   &  0.031  \\
\hline
$\sigma_{8}$  & 0.0072   &  0.011     &  0.0078    & 0.012   \\
\hline \hline
$w_1$  &  0.084  &  0.12     &  0.25     &   0.36 \\
\hline
$w_2$  &  0.074 &  0.11     &  0.31     &    0.48  \\
\hline
$w_3$  &  0.036  & 0.056      & 0.090     &    0.14   \\
\hline
$w_4$   & 0.020  &  0.031     & 0.029     &    0.046   \\
\hline
$w_5$  &  0.031  & 0.050      & 0.034     &  0.055  \\
\hline
$w_6$    & 0.41  &  0.52     & 0.53     &   1.5  \\
\hline
FOM    & 4700 &    1900   &    2200  &   870  \\
\hline
$\Omega_{c}$ & 0.025  &  0.041  & 0.048     &  0.080  \\
\hline
$\Omega_{b}$ & 0.016   & 0.029  &  0.020    &  0.037  \\
\hline
$h$  & 0.13   &   0.22    &  0.11     &  0.21  \\
\hline
$n$  & 0.030  &  0.047     &  0.027     &  0.047  \\
\hline
$\sigma_{8}$ & 0.018 &  0.027  & 0.058     &  0.095  \\
\end{tabular}
\caption{1$\sigma$ uncertainties on cosmological parameters
from WL alone, for the case of five tomographic bins in a flat universe without external priors. `S' stands for Smith et al., and `M' for MTC.}
\label{table4}
\end{center}
\end{table}

Table~\ref{table4} shows the uncertainties on cosmological parameters for three different
EOS parameterizations ($w_c$, $\{w_0,w_a\}$, and redshift-binned
$w(z)$), as determined from weak lensing with five tomographic bins. For a
redshift-binned $w(z)$, including the MTC corrections to the dark matter power
spectrum, an LSST weak lensing survey constrains five EOS parameters to better
than 10$\%$ (three of which are better than 5$\%$), and a JDEM-like survey
constrains three EOS parameters. The LSST binned FOM is a factor of 2.5 larger than
the one for JDEM, and the constraints on other cosmological parameters are
roughly a factor of two better. The main point here is that a very wide
survey is more effective than a deep, but narrow, survey. Although a deep,
moderately wide survey could be even more effective, the constraints depend
sensitively on the precise nature of the surveys.
The MTC improvement to the nonlinear dark matter power spectrum 
leads to a factor of two improvement in the FOM of a redshift-binned EOS. 

\begin{table}[t!]\footnotesize
\begin{center}
\begin{tabular}{r|cccccccccc|c}
\hline
No. of bins  & $\sigma (w_1)$ & $\sigma (w_2)$ & $\sigma (w_3)$& $\sigma (w_4)$ & $\sigma (w_5)$ & $\sigma (w_6)$ & FOM\\
\hline \hline
1 (M-LSST)     &   0.59  & 1.47  &  0.96 &  0.69 &  0.74 &  1.6 & 8.8    \\
\hline
2 (M-LSST)     & 0.14    & 0.14  & 0.074  & 0.041  & 0.046  & 0.30  & 1300   \\
\hline
3 (M-LSST)    &   0.099 &  0.085 & 0.043  &  0.024 & 0.032  & 0.53  & 3600   \\
\hline
4 (M-LSST)    &   0.082 &  0.075 & 0.038  &  0.020 & 0.031  & 0.47  & 4500   \\
\hline
5 (M-LSST)    &    0.084 & 0.074 & 0.036  & 0.020  & 0.031  & 0.41  & 4700   \\
$+ \Omega_k$    &  0.080 &  0.072 & 0.037  &  0.021 & 0.041  & 0.13  & 3900   \\
$+$ Priors    &    0.083 &  0.074 & 0.037  &  0.020 & 0.032  & 0.33  & 4600   \\
\hline
$l<1000$    &    0.12 & 0.10 & 0.052  & 0.029  & 0.050  & 0.16  & 2200   \\
$+ \Omega_k$    &  0.13 &  0.11 & 0.060  &  0.035 & 0.075  & 0.20  & 1400   \\
$+$ Priors    &    0.13 &  0.10 & 0.052  &  0.028 & 0.049  & 0.19  & 2200   \\
\hline
$z<1.5$    &    0.11 & 0.10 & 0.045  & 0.024  & 0.043  & 1.4  & 3000   \\
$+ \Omega_k$    &  0.20 &  0.18 & 0.088  &  0.052 & 0.40  & 0.52  & 560   \\
$+$ Priors    &    0.12 &  0.10 & 0.046  &  0.025 & 0.046  & 1.27  & 2700   \\
\hline
$+$ 2300 SNe    &    0.055 & 0.049 & 0.032  & 0.021  & 0.032  & 0.50  & 5000   \\
\hline
1 (S-LSST)      & 0.56   &   2.8 &  2.2 &  2.1 & 2.5  & 17  &  3.9  \\
\hline
2 (S-LSST)      &   0.29  & 0.90  & 0.25  & 0.065  & 0.043  & 0.17  & 850   \\
\hline
3 (S-LSST)     &    0.33 &  0.40 & 0.12  &  0.037 & 0.035  & 0.43  & 1700   \\
\hline
4 (S-LSST)     &    0.22 &  0.29 & 0.095  &  0.031 & 0.034  & 0.57  & 2100   \\
\hline
5 (S-LSST)    &    0.25 &  0.31 & 0.090  &  0.029 & 0.034  & 0.53  & 2200   \\
$z<1.5$    &    0.26 & 0.28 & 0.087  & 0.032  & 0.047  & 1.0  & 1600   \\
\hline
2 (M-JDEM)     &   0.22 & 0.22  & 0.12  & 0.067  & 0.082  & 0.52  & 490   \\
\hline
5 (M-JDEM)    &     0.12 & 0.11  &  0.056 &  0.031 &  0.050 & 0.52  & 1900   \\
$+ \Omega_k$    &  0.12 &  0.11 & 0.056  &  0.032 & 0.062  & 0.22  & 1700   \\
$+$ Priors    &    0.12 &  0.11 & 0.056  &  0.031 & 0.051  & 0.49  & 1900   \\
\hline
$l<1000$    &    0.20 & 0.17 & 0.086  & 0.048  & 0.085  & 0.26  & 790   \\
$+ \Omega_k$    &  0.22 &  0.18 & 0.097  &  0.056 & 0.12  & 0.32  & 550   \\
$+$ Priors    &    0.21 &  0.17 & 0.086  &  0.047 & 0.081  & 0.30  & 810   \\
\hline
$z<1.5$    &    0.18 &  0.16 & 0.073  &  0.040 & 0.078  & 1.7  & 1100  \\
$+ \Omega_k$   & 0.27 & 0.26 & 0.13  & 0.075  & 0.67  & 0.75  & 270   \\
$+$ Priors    &    0.18 &  0.16 & 0.073  &  0.040 & 0.079  & 1.6  & 1000   \\
\hline
$+$ 2300 SNe    &    0.067 & 0.061 & 0.045  & 0.033  & 0.056  & 0.58  & 2200   \\
\hline
2 (S-JDEM)      &   0.45  & 1.3  & 0.43  & 0.11  & 0.075  & 0.29  & 280    \\
\hline
5 (S-JDEM)    &     0.36 &  0.48  & 0.14  & 0.046  & 0.055  & 1.5 & 870   \\
$z<1.5$    &    0.36 & 0.45 & 0.14  & 0.052  & 0.082  & 1.5  & 580   \\
\end{tabular}
\caption{1$\sigma$ constraints on values of $w(z)$ in uncorrelated redshift
bins, where `S' stands for Smith et al., and `M' for MTC. The EOS is redshift
binned at $z = \{ 0.07, 0.15, 0.3, 0.6, 1.2, 3.0\}$, but decorrelating
the covariance matrix for the original bins results in a leakage across bins.
Figure~\ref{figure8} shows this leakage in terms of window functions. The 2300 SNe alone constrain the
six EOS bins to [0.11,0.10,0.12,0.69,0.24,1.2].}
\label{table5}
\end{center}
\end{table}

It is possible to compare our weak lensing constraints with the combined constraints from SNe,
BAOs, and CMB. The latter have been calculated in Sarkar et al.~\cite{Sarkar},
using a Markov Chain Monte Carlo (MCMC) likelihood approach.
Case C in Sarkar et al. considers a mock catalog of 300 SNe at $z<0.1$~\cite{Aldering2002} and 2000 SNe in the range of 
$0.1<z<1.8$~\cite{Aldering}, two current BAO distance estimates~\cite{EisS, PercivalS}, fifteen optimistic future BAO measurements (five from V1N1
of~\cite{SeoS}, ten from ADEPT~\cite{RiessS}), as well as CMB constraints on
$[\Omega_m h, h, R]$ of [0.023, 0.08, 0.03]~\cite{Freedman, TegmarkS, WangS},
where $R$ is the distance to the last scattering surface. For this extensive
data set, the redshift-binned dark energy FOM is 1300. In comparison, our projected
lensing constraints yield an FOM of 4700 for LSST and 1900 for
JDEM (for the MTC matter power spectrum). It is evident that weak lensing 
is potentially an extraordinarily powerful dark energy probe.

\begin{figure}[!b]
\epsfxsize=3.4in
\centerline{\epsfbox{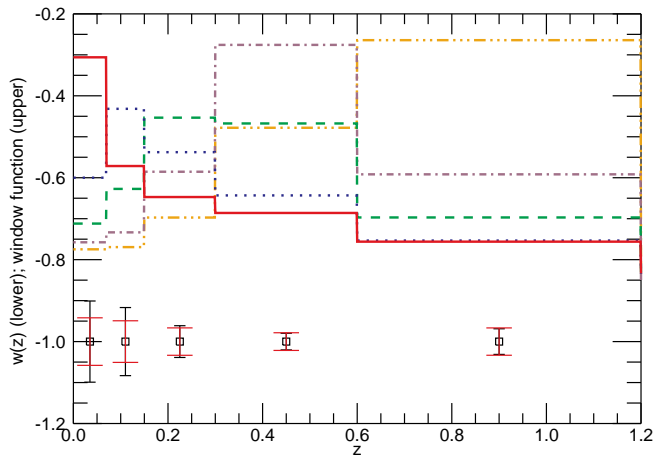}}
\caption{Lower portion of the figure shows the constraint on
the dark energy EOS in five redshift bins. The narrow (black) error bars
are obtained from weak lensing alone, for five tomographic bins, with an
HST prior on the Hubble constant. The wide (red) error bars are obtained from a joint
analysis of WL and SNe. Note that we are neglecting WL systematic uncertainties in this plot.
The upper half of the figure illustrates the window functions associated with the first five decorrelated dark
energy bins (WL + SNe). (Note that the window functions have been shifted down by
a constant of 0.8 for visual clarity.)}
\label{figure8}
\end{figure}

\subsection{SN/Weak Lensing Complementarity}

Our distribution of lensing source galaxies peaks at a redshift of unity. It is 
therefore interesting to combine our higher-redshift lensing measurements with
those from supernovae. To this end, we uniformly distribute a set of 300 SNe at
$z<0.1$~\cite{Aldering2002}, and 2000 SNe in the range $0.1<z<1.8$ (as expected
from a space-based JDEM probe~\cite{Aldering}, or as part of a first data
release by a ground-based telescope such as LSST). For each supernova we take
the intrinsic noise to be a Gaussian in magnitude with $\sigma_{\rm int} =
0.1$~\cite{KimS}. We divide the Hubble diagram for $z>0.1$ into 50 redshift
bins, and associate each bin with a redshift-dependent systematic floor of
magnitude $ \delta_m = 0.02(0.1/{\Delta z})^{1/2}(1.7/z_{\rm max})(1+z)/2.7$~\cite{LinderS}. We assume 
that this irreducible systematic has no correlation
between bins. For the SN constraints we use an HST prior of 0.08 for $h$~\cite{Freedman} (which is 
also applied to the WL case in the comparison with SNe+WL). For further details on the
SN approach we refer the reader to earlier work~\cite{Sarkar}.

\begin{table}[!t]\footnotesize
\begin{center}
\begin{tabular}{r|cccccccccc|c}
\hline
Probe  & $| \delta (w_0) |$ & $| \delta (w_a) |$ & FOB($w_0,w_a$)& $| \delta (w_c) |$ & FOB($w_c$)\\
\hline
LSST   &  0.073       & 0.19   & 1.9     &  0.026 & 1.8 \\
\hline
JDEM   &  0.076       & 0.20   & 1.3     &  0.027 & 1.2 \\
\end{tabular}
\begin{tabular}{r|cccccccccc|c}
\hline
Probe  & $| \delta (w_1) |$ & $| \delta (w_2) |$ & $| \delta (w_3) |$& $| \delta (w_4) |$ & $| \delta (w_5) |$ & $| \delta (w_6) |$\\
\hline
LSST    & 0.0050       & 0.060       & 0.080       & 0.034        & 0.0026       & 0.029\\
\hline
JDEM    & 0.0039       & 0.20       & 0.18       & 0.058        & 0.0069       & 0.031\\
\end{tabular}
\caption{Bias on the determination of the dark energy
equation of state, for the case of five tomographic bins (in a flat universe with no external
priors). For the FOB of $\{w_0,w_a \}$, a value of 1.5 indicates a 1$\sigma$
shift from the true estimate, and a value of 2.5 indicates a 2$\sigma$ shift~\cite{Press, Shapiro}. For $w_c$, these 
values are 1.0 and 2.0, respectively. The lower entries list the bias on the {\it correlated}\/
$w(z)$ bins.}
\label{table6}
\end{center}
\end{table}

The uncorrelated redshift-bin uncertainties in five tomographic weak lensing
bins are given in Tables~\ref{table4} and~\ref{table5}, and illustrated in Figure~\ref{figure8} for an LSST data set. The 
narrow (black) error bars are the constraints from weak lensing alone, while the wide (red) error bars represent
the constraints from both weak lensing and SNe. Although weak lensing alone
constrains three redshift bins to 5$\%$, by combining this data set with SNe we
constrain two additional EOS parameters to that level. Weak lensing and SNe are
complementary, as weak lensing is most effective for constraints at intermediate
redshifts (roughly $0.2<z<1.2$), while future SN data provide a more effective
probe at lower redshifts ($z<0.2$).

\subsection{Bias in dark energy due to uncertainties in the nonlinear matter power spectrum}

We have calculated the improvement in weak lensing constraints from 
use of the MTC matter power spectrum (incorporating $w(z) \neq -1$). It is
also interesting to calculate the bias that would arise from use of the more approximate
Smith et al.~power spectrum, when the real data is described by MTC. The bias in each
parameter is given by~\cite{Knoxetal, Shapiro}:
\begin{equation}
\delta p_{\alpha} = f_{\rm sky} \sum_{l, \beta} F^{-1}_{\alpha \beta}
{(2l+1)\Delta l \over 2} \Tr \left [ \tilde{{\bf C}}_l^{-1} {{\partial {\bf
C}_l} \over {\partial p_\beta}} \tilde{{\bf C}}_l^{-1} \delta C_l \right ],
\end{equation}
where $\delta C_l$ is the difference between the MTC and Smith et al.~convergence
spectra. One can calculate the corresponding
figure of bias (FOB) for the subset of EOS parameters as rendered by~\cite{Shapiro}:
\begin{equation}
{\rm FOB} = \left ( \sum_{\alpha , \beta} \delta p_{\alpha} \tilde{{F}}_{\alpha
\beta} \delta p_{\beta} \right )^{1/2}.
\end{equation}
Thus, for the case of one parameter, the FOB is simply equal to the ratio of
that parameter's bias to its uncertainty. Table~\ref{table6} presents the FOB
values for LSST and JDEM weak lensing measurements. We find a
1$\sigma$--2$\sigma$ bias
between the Smith et al.~and MTC determinations of the dark energy EOS (for a $w=-0.9$
fiducial cosmology). {\it It is therefore critical that the nonlinear matter power
spectrum be well-characterized, to enable precision constraints on a dynamical dark
energy equation of state.}

\subsection{Systematic Uncertainties}

\begin{figure}[!t]
\epsfxsize=3.4in
\centerline{\epsfbox{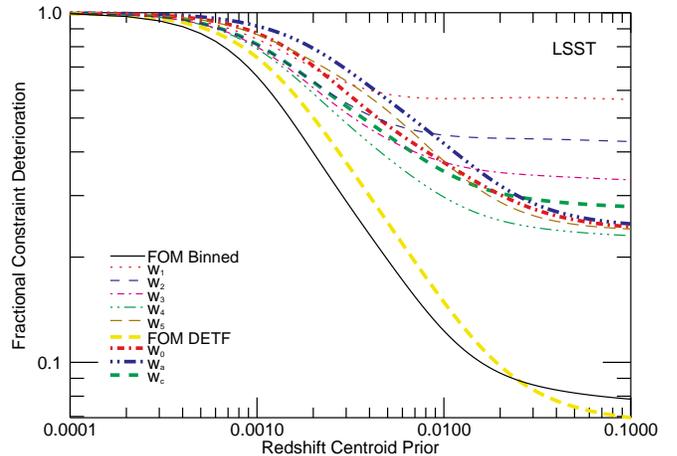}}
\caption{Deterioration of constraints as a function of a Gaussian prior on each centroid
of the redshift bins for LSST. We plot both
$\rm{FOM_{syst+prior}/FOM_{no-syst}}$ and $\sigma(w)|_{\rm{no-syst}}/\sigma(w)|_{\rm{syst+prior}}$.}
\label{figure9}
\end{figure}

\begin{figure}[!t]
\epsfxsize=3.4in
\centerline{\epsfbox{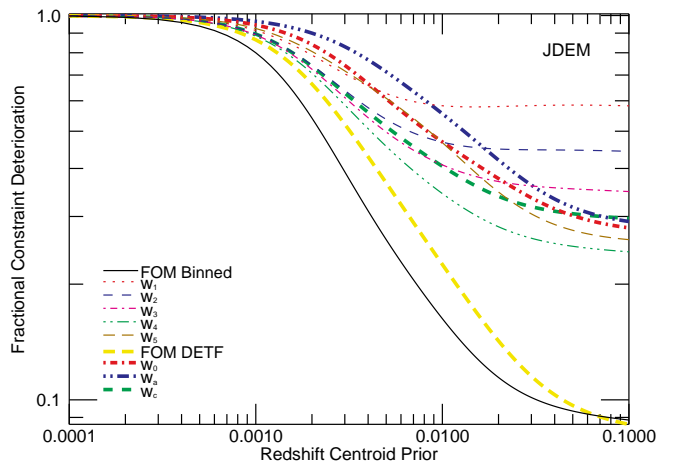}}
\caption{Deterioration of constraints as a function of the prior on the centroid of the redshift bins for a JDEM survey.}
\label{figure10}
\end{figure}

To assess the importance of systematic errors in the weak lensing measurements, we
approximate a possible shear miscalibration by including a multiplicative
factor, $f_i$, for each tomographic bin. 
We ignore additive systematic terms to the convergence power spectrum, as their dependence 
on $l$ is currently poorly understood~\cite{Hannestad}.
The uncertainty in the redshift distribution of the sources is modeled to first
order as a shift in the centroid of the tomographic bins~\cite{Hut06}.  We
ignore other possible sources of systematics, such as baryonic
uncertainties in the nonlinear scales of the matter power spectrum~\cite{Dragan}, and
intrinsic ellipticity correlations of the source galaxies~\cite{Hirata}.

\begin{figure}[!t]
\epsfxsize=3.4in
\centerline{\epsfbox{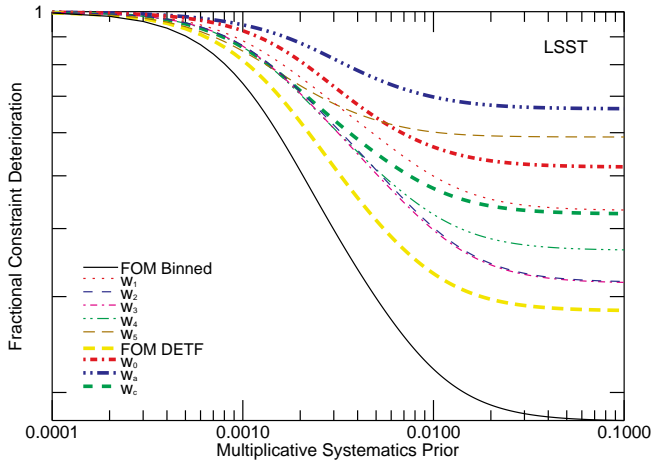}}
\caption{Deterioration of constraints as a function of the Gaussian prior on each multiplicative factor in shear for LSST.}
\label{figure11}
\end{figure}

\begin{figure}[!t]
\epsfxsize=3.4in
\centerline{\epsfbox{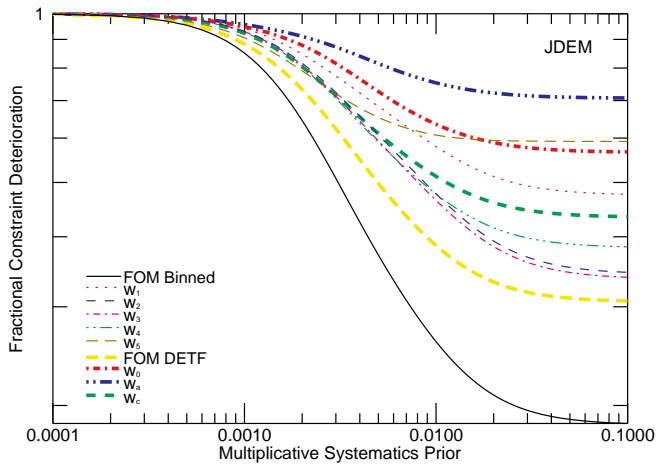}}
\caption{Deterioration of constraints as a function of the prior on multiplicative factors in shear for a JDEM survey.}
\label{figure12}
\end{figure}

Following Huterer et al.~(2006)~\cite{Hut06}, the cross-correlation terms are
changed by the systematics as: $C_{ij}(l) = C_{ij} \left( l; ~ z_i + \delta
z_i, z_j + \delta z_j \right) \left[ 1 + f_i + f_j \right]$. For our
five tomographic bins we therefore introduce an additional $5 \times 2 = 10$
free parameters to encapsulate possible systematics. The effect of
these systematic uncertainties on the measurement of dark energy 
is heavily influenced by the corresponding priors. If we are optimistic, and
claim an understanding of the systematics at the 0.1\% level, the deterioration
in dark energy constraints is essentially negligible. However, the constraints
are strongly compromised if the systematics are unknown 
at the level of 1\%. This is illustrated in Figures 9--12.

Both of the surveys offer the potential for self-calibration tests, if the priors
are above the percent level. For LSST, there is a factor of five (three) degradation
of the binned FOM (DETF FOM) for multiplicative shear errors and a factor of 14
deterioration in the two FOMs for errors in the redshift bin centroids. In
general the degradations are slightly milder for JDEM; e.g. a factor of 11
degradation in the two FOMs for redshift bin centroid
uncertainties. We note that a combined analysis with the lensing bispectrum has
the potential to improve these prospects~\cite{Hut06}.

\section{Conclusions}

We calculate weak lensing constraints on the dark energy equation of state,
incorporating an improved nonlinear matter power spectrum which accounts for the
effects of a dynamical dark energy. The most commonly utilized nonlinear extension of
the matter power spectrum, calibrated from N-body simulations by Smith et al.,
does not incorporate the effects of time-evolving dark energy. We follow a
prescription presented by McDonald, Trac, $\&$ Contaldi (2006), and extend
the Smith et al.~form to include evolving dark energy. We then perform a full
Fisher matrix analysis for two prospective weak lensing surveys, utilizing the improved power spectrum.

By considering weak lensing tomography with an improved nonlinear matter power spectrum
that incorporates dynamical dark energy, we find a factor of two improvement in the
dark energy figure of merit (for the WMAP5 fiducial cosmology). Although the changes in the
matter power spectrum and the lensing kernel somewhat cancel in their effects on
the convergence power spectrum, the parameter correlations nonetheless lead to
improved parameter constraints.

We further show that a poor approximation (off by $\gsim10\%$) to the
dark energy corrections of the nonlinear matter power spectrum leads to a $\gsim1\sigma$ bias on the dark
energy equation of state. Future weak lensing surveys must
therefore incorporate percent-level accurate dark energy modifications to the nonlinear matter power spectrum to 
avoid introducing significant bias in their measurements.
In addition, by combining weak lensing data with supernova
measurements at lower redshifts, we show that a general dark energy model can be constrained in five
redshift bins to 5\% for an LSST-type survey and to 10\% for a JDEM-like probe. These dark energy constraints are contingent
upon our ability to understand weak lensing systematic uncertainties, such as those
arising from shear miscalibration and the redshift uncertainty of the sources.
If it is possible to control these systematics to 0.1\%, weak lensing
constraints of dynamical dark energy from next-generation surveys offer
tremendous promise.

\smallskip
{\it Acknowledgments:} We are grateful to Alexandre Amblard and Paolo Serra for helpful conversations throughout this work. 
We also thank Dragan Huterer, Patrick McDonald, Charles Shapiro, and Licia Verde for useful discussions.
S.J. acknowledges support from a GAANN fellowship at UCI. This work was also
supported by NSF CAREER AST-0645427 (A.C.) and LANL IGPP Astro-1603-07 (D.E.H.).

\end{document}